\documentclass[preprint,showpacs,showkeys,byrevtex,superscriptaddress]{revtex4}
\usepackage{graphics}
\begin{document}
\newenvironment{tab}[1]
{\begin{tabular}{|#1|}\hline}
{\hline\end{tabular}}

\title {
Detection efficiency of a BEGe detector using the Monte Carlo method and a comparison to other calibration methods
}

\author{N. Stefanakis}
\affiliation{
GMA Gamma measurements and analyses e.K. 
PO Box 1611, 72706 Reutlingen, Germany
}  
\date{\today}

\begin{abstract}
In this paper we model by using the Monte Carlo simulation code PENELOPE 
\cite{salvat,lepy} a Broad Energy Germanium (BEGe) detector and determine its efficiency. The simulated geometry consists of a point source located close to the detector as well as volume sources with cylindrical geometry. A comparison of the simulation is made to experimental results as well as to analytical calculations.
\end{abstract}
\maketitle

\section{Introduction}

Gamma-ray spectroscopy is one of the most used methods for the characterisation of nuclear waste, because of its non destructive character without the need of sample preparation. Usually the necessary efficiency calibration is done experimentally using standard radioactive sources of the same geometry, density and chemical composition as the sample to be measured. This procedure is not very effective.

One effective procedure to overcome these difficulties is the use of Monte Carlo simulation. It can be defined as a methodology for obtaining estimates of the solution of mathematical problems by means of random numbers. In the efficiency calculation the history of the primary particle namely the photon and of the secondary particles resulting from the interaction in the detector is tracked until all its energy is dissipated. Monte Carlo method allows the peak efficiencies to be calculated taking into account the characteristics of the detector and the measurement geometry. 

For example, one generalist Monte Carlo simulation code is the MCNP \cite{briesmeister} used to simulate neutrons, electrons, photons interactions. The user creates an input file that is subsequently read by MCNP. This file contains information about the problem in areas such as: the geometry specification, the description of materials and selection of cross-sections evaluations, the location and characteristics of the neutron, photon, or electron source, the type of answers or tallies desired, and any variance reduction techniques used to improve efficiency. 

One user friendly approach based on Monte Carlo analysis is the “In Situ Object Counting System (ISOCS)” from the company CANBERRA \cite{isocs}. In that case the measurement parameters are entered in a geometry editor called Geometry Composer using mathematical templates for most of the commonly encountered geometries. Then the efficiency of the measurement is calculated. This approach supposes that the detector is characterised; A procedure which is time- an cost-expensive. The user is also limited to the available geometrical templates. 

In this paper we apply the Monte Carlo code PENELOPE, developed to simulate the interaction of photons, electrons and positrons with matter, in order to calculate the efficiency calibration of a BEGe detector. On PENELOPE the existing geometry package
 permits the generation of random electron-photon showers in
 material systems consisting of homogeneous bodies limited by quadric surfaces, i.e.,
 planes, spheres, cylinders, etc. \cite{salvat}.
The simulated geometry considered in this paper is a Plutonium point source located close to the detector. 
Other more realistic geometries e.g. the cylindrical geometry, as well as 
comparison to analytical calculations are presented.

\section{Results and discussion}

\subsection{Point source}

\subsubsection{Measurement geometry}

\begin{figure}
\scalebox{0.6}{\includegraphics{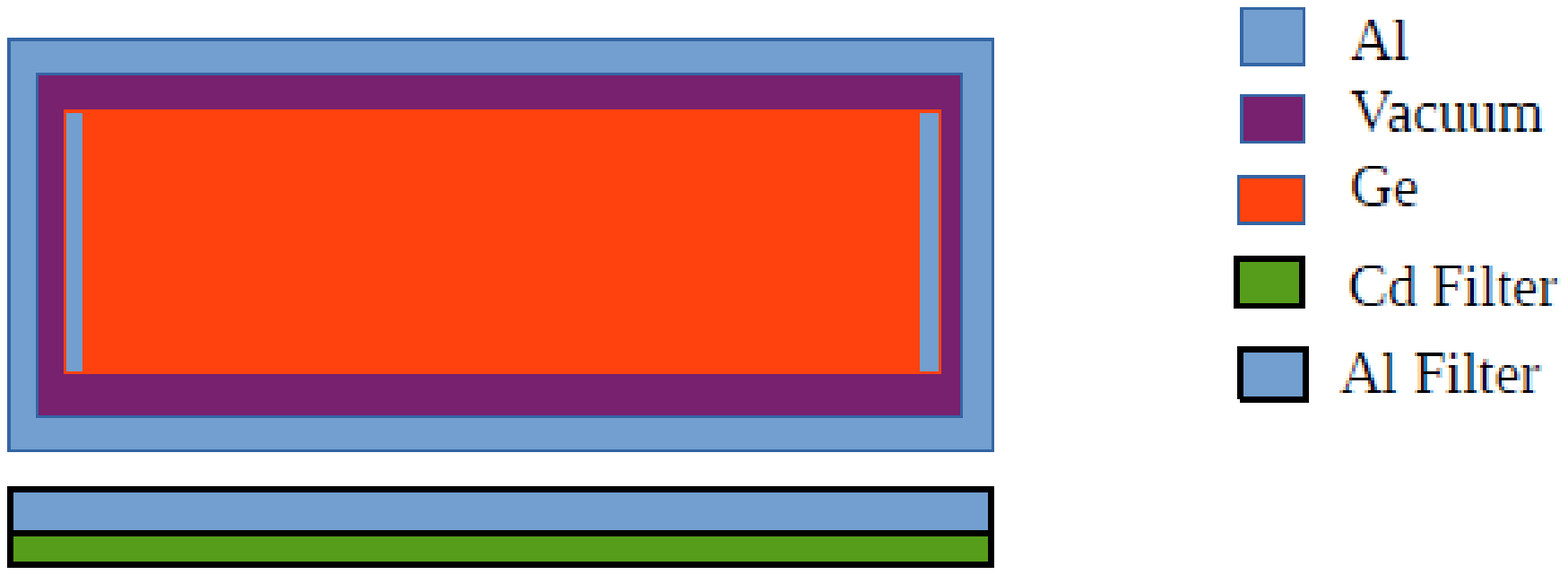}}
\caption{Measurement geometry}
\vspace{2cm}
\label{pointsource.fig}
\end{figure}

The detector was simulated as a simple cylinder made of Germanium (Ge) of radius $35$mm
and height $25$mm hold by an aluminium (Al) cylinder of width $1$mm. The detector core is
wrapped by an aluminium case of width $2$mm. The space between the aluminium holder and
the aluminium case is vacuum ($2-3$mm). The high quality vacuum is necessary to minimize
the collection of contaminations on the detector surfaces which cause degradation of the
performance. It also deans as a thermal shield to prevent the transfer of heat from the outer
surfaces to the detector. A material filter made of Aluminium $4$mm and Cadmium $0.5$mm is
placed in frond of the detector. The filter is used in order to attenuate the low energy gamma
radiation. In Fig. \ref{pointsource.fig} we plot a two dimensional xz cross section of the detector geometry. The
detector parameters have been chosen to simulate a BEGe detector and are seen in Table \ref{table1}.
\begin{table}
\scalebox{0.8}{\begin{tabular}{|c|c|}
\hline
{\bf Detector Parameters} & {\bf Dimension (mm)}\\
\hline
Ge crystal diameter & 70\\
\hline
Ge crystal height & 25\\
\hline
Al holder & 1\\
\hline
Vacuum layer & 3\\
\hline
Al case & 2\\
\hline
Al Filter & 4\\
\hline
Cd Filter & 0.5\\
\hline
\end{tabular}
}
\caption{Detector parameters} 
\label{table1}
\end{table}

Several types of electrical contacts are used in order to apply the electric field such as Li diffused
layer for one contact and an evaporated or ion-implanted metal film for the other.
They produce a layer of attenuating material, from which the charge carriers are not collected
and therefore called dead layers. For the sake of simplicity the Germanium dead layer was not
simulated. The existence of a Germanium dead layer would have decreased the active detector
volume and also attenuated the low energy gamma rays.
The Pu-239 point source is located $66$cm from the detector window. The incident beam
photons move along the z-axis (i.e. upwards) and impinge normally on the surface of the
detector.

\subsubsection{simulated spectra}
\begin{figure}
\centering\scalebox{0.6}{\includegraphics{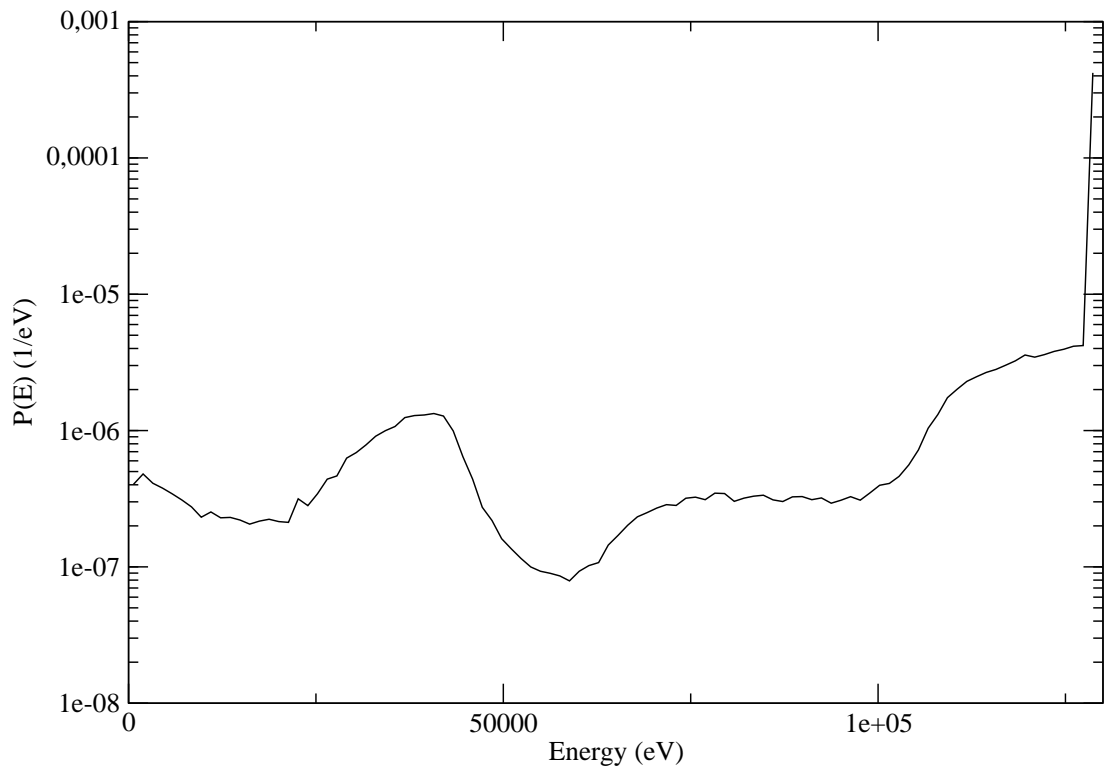}}
\caption{Distribution of energy deposited in the Ge crystal} 
\vspace{2cm}
\label{distribution.fig}
\end{figure}

Using Monte Carlo code a narrow beam of energy $129$ KeV \cite{reilly,mini,debertin} corresponding to a gamma
ray from Pu-239 enters the detector. The measurement time is $300$ sec. $10^6$ particles were
simulated.
We present in Fig. \ref{distribution.fig} the distribution of the deposited energy in the Germanium crystal $P(E)$
$(eV^{-1})$ as a function of the energy. This plot can be considered equivalent to the simulated
spectra. The photopeak at the energy $129$KeV is seen as well as the Compton edge.

\subsubsection{measurement efficiency}

\begin{figure}
\scalebox{0.6}{\includegraphics{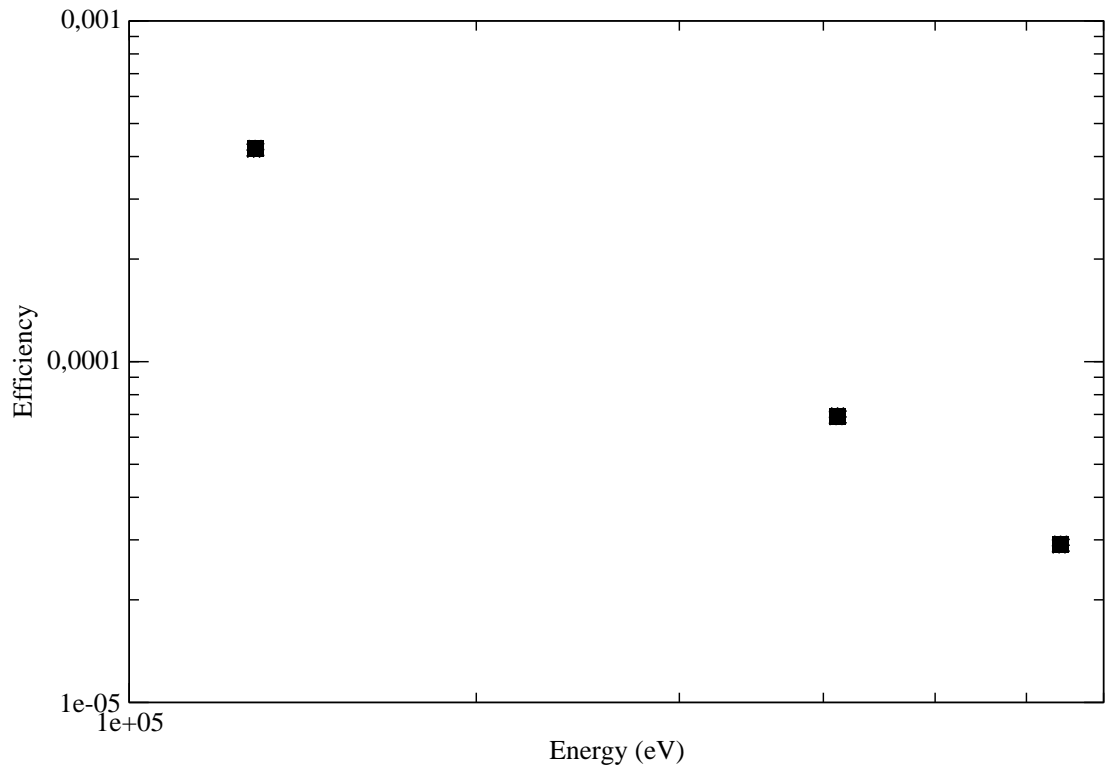}}
\caption{Detection efficiency in the range 0 bis 645KeV} 
\label{efficiencyAlHolderCd.fig}
\end{figure}

The count rates or areas of individual peaks in the spectrum are related to the amount of
radioactive material deposited in the body by a factor of efficiency. This efficiency depends
on several factors including the intrinsic efficiency of the detector, the geometry of the
measurement (solid angle, position, shape of the individual to be measured and the
radionuclide distribution in the body) and the properties of the nuclide.
Experimentally the detection efficiency $Eff_{E_{\gamma}}$ at energy $E_{\gamma}$ is determined as follows:
\begin{equation}
Eff_{E_{\gamma}}=\frac{N_{E_i}}{Amtf}
\end{equation}
where $N_{E_i}$ is the net area under the full energy peak corresponding to energy photons $E_{\gamma}$
emitted by a radionuclide with known activity $A$, $f$ is the emission probability, $m$ is the sample
mass and $t$ is the counting time. Usually calibration methods of multienergy gamma ray
emmiters like Eu-152 are used. This nuclide has half life of $13.5$ years and it covers the energy
range from $122$ KeV to $1408$ KeV. The disadvantages are the Compton edges and the coincident
summing corrections associated with each photopeak. As a next step an interpolation is
performed in order to obtain the efficiency for every energy value.
One of the output of the Monte Carlo code is the probability distribution of the energy
deposited in the Germanium crystal. From this distribution the peak efficiency values (total
absorption detection efficiency) can be directly calculated.
However the number of counts defined as belonging to the measured full-energy peak need
not correspond to the number of full energy absorption events calculated by the monte carlo
code. In particular the low energy tail of a measured peak may be only partly included in the
experimental peak area, but all of these events are included in the calculated peak area.
Therefore the experimental results are only qualitatively compared to the simulated ones.
The full energy peak efficiency of the detector as a function of the energy is presented in Fig. \ref{efficiencyAlHolderCd.fig}.
The calculated values for the efficiency are compared to the values of the efficiency of a
BEGe detector using the ISOCS calibration software from Camberra, as well as to the
detection efficiency calculated using the internet portal nucleonica for the same measurement
conditions (see \cite{stefanakis}). The deviation may be attributed to the fact that the detector parameters
are not accurately known and therefore a direct comparison is not possible.

\subsection{Volume sources}
\subsubsection{Semi empirical method}

Having calculated or measured the efficiencies for simple geometries e.g. point sources one
can use semi-empirical methods in order to derive the efficiency for other complicated
geometries e.g. disc or cylinder.
Let us consider a cylindrical homogeneous source with its symmetry axis coinciding with the
detector axis. The detection efficiency of this structure can easily be calculated numerically.
We can consider this source as a multilayer disk source with each layer at a distance $x$ from
the detector. The efficiency is calculated as \cite{debertin}:
\begin{equation}
\epsilon(d)=\frac{2}{R^2}\frac{1}{h}\int_d^{d+h}\int_0^{R}\epsilon_P(r,x)rdrdx
\end{equation}
where $h,R$ is the source height and radius respectively and $d$ denotes the distance to the source
end nearest to the detector. $\epsilon_P(r,x)$ is the radial dependence of the point source efficiency
(area efficiency) given for example by
\begin{equation}
\epsilon_P(r,x)=\epsilon_P(0,x)\frac{x^2}{x^2+r^2}
\end{equation}
$\epsilon_P(0,x)$ is the point efficiency at the detector axis at distance x from the detector.
\begin{equation}
\epsilon_P(0,x)=\frac{1}{2}[1-(1+\frac{R_D^2}{x^2})^{-1/2}-
\frac{3}{8}R^2\frac{R_D^2}{x^4}(1+\frac{R_D^2}{x^2})^{-5/2}-...]
\end{equation}
where $R_D$ is the detector radius.

\subsubsection{Numerical method}
\begin{figure}
\scalebox{0.6}{\includegraphics{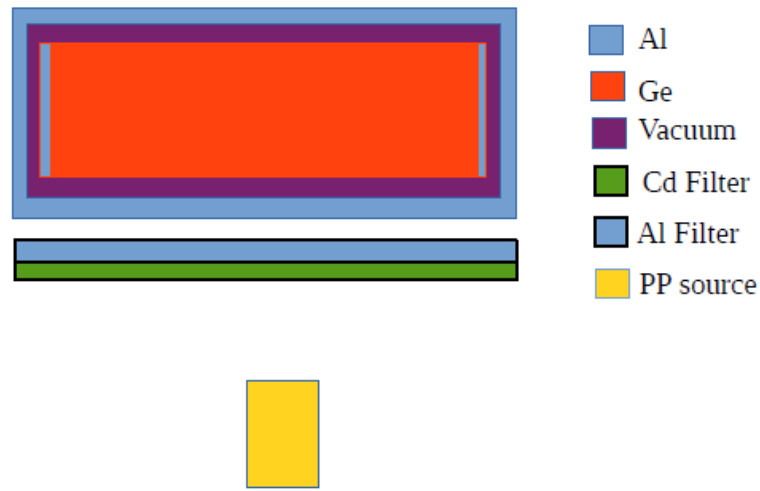}}
\caption{Measurement geometry for the volume source} 
\vspace{2cm}
\label{volumesource.fig}
\end{figure}

\begin{figure}
\scalebox{0.6}{\includegraphics{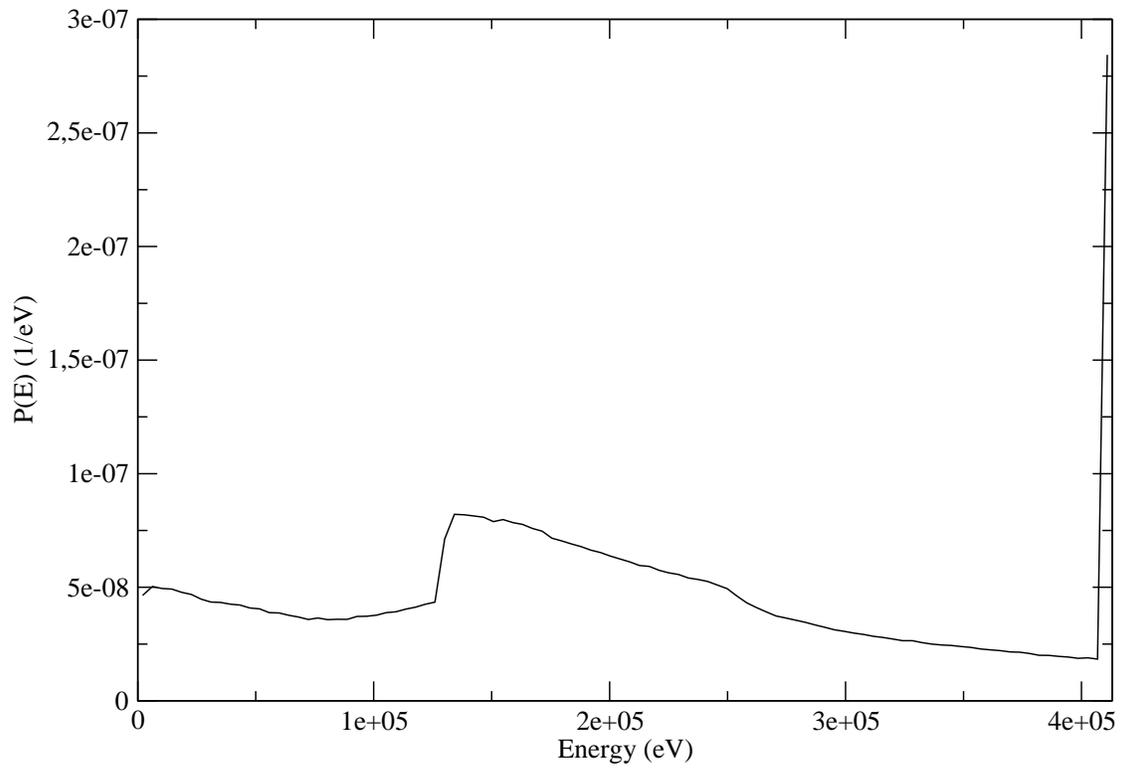}}
\caption{Distribution of energy deposited in the Ge crystal} 
\vspace{2cm}
\label{distributioncyl.fig}
\end{figure}

We are going to study the effect of the volume of the source to the simulated spectra as well
as in the efficiency.
We consider the measuring geometry of Fig. \ref{volumesource.fig}. The detector parameters are the same as in
section II where a point source was studied. Here the source is cylindrical of height $1$cm and
radius $1$cm, positioned $10$cm apart from the frond edge of the detector. A uniform activity
concentration was assumed. We consider the Nuclide Plutonium 239.
The simulated spectra is presented in Fig. \ref{distributioncyl.fig}.
For simplicity only two gamma energies of the
isotope having large emission probability were studied.

\begin{figure}
\scalebox{0.6}{\includegraphics{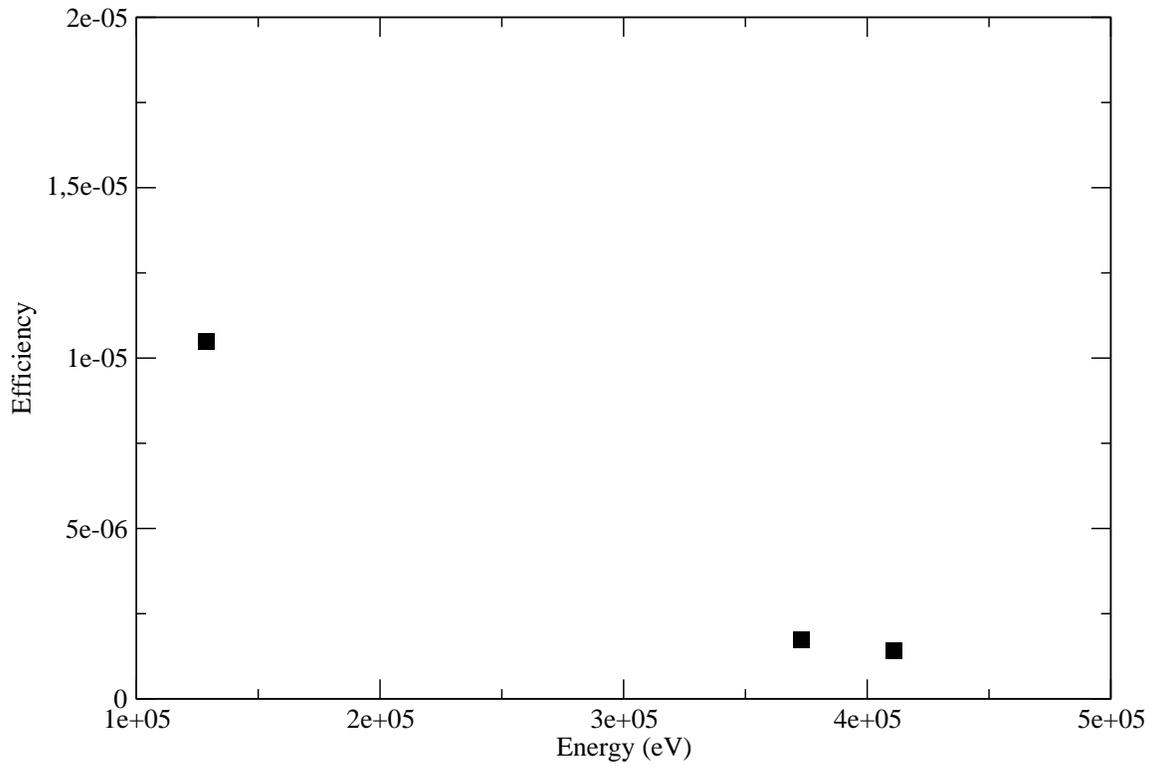}}
\caption{Detection efficiency for the volume source geometry} 
\vspace{2cm}
\label{volumesourceefficiency.fig}
\end{figure}

It is seen that the gamma peak at 129KeV is affected from the Compton background created
from the peak at $413$KeV.
We plot in Fig. \ref{volumesourceefficiency.fig} the efficiency of the peak measurement as a function of the energy.

\section{conclusions}
We studied numerically the efficiency of a BEGe detector in the point source as well as in the
volume source geometry. The results are well compared to experimental measurements. The
efficiency calculated with the monte carlo method is comparable to the efficiency obtained by
other methods e.g. use of calibration standards and can be directly integrated to the
measurement procedure.

\bibliographystyle{prsty}


\newpage

\end{document}